\documentclass[aps,prl,onecolumn,showpacs,amsmath,amssymb,floatfix,footinbib,superscriptaddress]{revtex4-1}
\usepackage{amsmath,graphicx,subfigure,amssymb,graphics,amsmath,mathrsfs,CJK,color,natbib}
\usepackage{multirow,fancyhdr,color,bm,tabularx,psfrag,dcolumn}
\usepackage[colorlinks=true,linkcolor=blue,urlcolor=blue,citecolor=blue]{hyperref}
\usepackage{tikz,xcolor,hyperref}
\definecolor{lime}{HTML}{A6CE39}
\DeclareRobustCommand{\orcidicon}{%
    \begin{tikzpicture}
    \draw[lime, fill=lime] (0,0)
    circle [radius=0.16] node[white]
   {{\fontfamily{qag}\selectfont \tiny ID}};\draw[white, fill=white] (-0.0625,0.095)
    circle [radius=0.007];
    \end{tikzpicture}
    \hspace{-2mm}}
\foreach \x in {A, ..., Z}
{\expandafter\xdef\csname orcid\x\endcsname{\noexpand\href{https://orcid.org/\csname orcidauthor\x\endcsname}{\noexpand\orcidicon}}}

\begin{document}

\title{\large Delayed response to the photovoltaic performance in a double quantum dot photocell with spatially correlated fluctuation }

\author{Sheng-Nan Zhu }
\affiliation{Center for Quantum Materials and Computational Condensed Matter Physics, Faculty of Science, Kunming University \\of Science and Technology, Kunming, 650500, PR China}
\affiliation{Department of Physics, Faculty of Science, Kunming University of Science and Technology, Kunming, 650500, PR China}

\author{Shun-Cai Zhao\textsuperscript{\orcidA{}}}
\email[Corresponding author: ]{zsczhao@126.com}
\affiliation{Center for Quantum Materials and Computational Condensed Matter Physics, Faculty of Science, Kunming University \\of Science and Technology, Kunming, 650500, PR China}
\affiliation{Department of Physics, Faculty of Science, Kunming University of Science and Technology, Kunming, 650500, PR China}

\author{Lu-Xin Xu }
\affiliation{Center for Quantum Materials and Computational Condensed Matter Physics, Faculty of Science, Kunming University \\of Science and Technology, Kunming, 650500, PR China}
\affiliation{Department of Physics, Faculty of Science, Kunming University of Science and Technology, Kunming, 650500, PR China}
\author{Lin-Jie Chen }
\affiliation{Center for Quantum Materials and Computational Condensed Matter Physics, Faculty of Science, Kunming University \\of Science and Technology, Kunming, 650500, PR China}
\affiliation{Department of Physics, Faculty of Science, Kunming University of Science and Technology, Kunming, 650500, PR China}

\begin{abstract}
A viable strategy for enhancing photovoltaic performance in a double quantum dot (DQD) photocell is to comprehend the underlying quantum physical regime of charge transfer. This work explores the photovoltaic performance dependent spatially correlated fluctuation in a DQD photocell. A suggested DQD photocell model was used to examine the effects of spatially correlated variation on charge transfer and output photovoltaic efficiency. The charge transfer process and the process of reaching peak solar efficiency were both significantly delayed as a result of the spatially correlated fluctuation, and the anti-spatial correlation fluctuation also resulted in lower output photovoltaic efficiency. Further results revealed that some structural parameters, such as gap difference and tunneling coefficient within two dots, could suppress the delayed response, and a natural adjustment feature was demonstrated on the delayed response in this DQD photocell model.  Subsequent investigation verified that the delayed response was caused by the spatial correlation fluctuation, which slowed the generative process of noise-induced coherence, which had previously been proven to improve quantum photovoltaic performance in quantum photocells. While anti-spatial correlation fluctuation and a hotter thermal ambient environment could diminish the condition for noise-induced coherence, as demonstrated by the reduced photovoltaic capabilities in this suggested DQD photocell model. As a result, we expect that regulated noise-induced coherence, via spatially correlated fluctuation, will have a major impact on photovoltaic qualities in a DQD photocell system. The discovery of its underlying physical regime of quantum fluctuation will broaden and deepen understanding of quantum features of electron transfer, as well as provide some indications concerning quantum techniques for high efficiency DQD solar cells.

\begin{description}
\item[Keywords]{Delayed response; spatially correlated fluctuation; double quantum dots photocell }
\item[PACS ]{73.23.-b, 73.63.Kv, 42.50.Hz }
\end{description}
\end{abstract}

\maketitle
\section{Introduction}

Recently, in many experimental works, the spatially correlated fluctuations between the ambient environment and neighboring chromophores were identified as the underlying regime\cite{2007Evidence777,2010Efficient2,2010Coherent11,2010Long111, 2010Coherently333,2014Combined,2017Signatures}  of quantum coherent energy transfer, and it was also proposed to be capable of protecting long-lived quantum coherence with the strong system-bath coupling condition, so it was regarded as the source of coherence beatings in light-harvesting complexes\cite{2017Role,2021Coherentw}. In Ref.\cite{PhysRevE.78.050902}, a unified theory was presented to account for spatially correlated conformational a static variations and high-frequency a dynamic fluctuations in densely packed pigment-protein complexes, and the calculated photon-echo signals demonstrated long-lasting coherence, which was also observed in experiments\cite{PhysRevE.78.050902}. Other researchers distinguished between the energy transfer process, quantum coherence signatures, and line shape of an exciton with spatially correlated and uncorrelated bath\cite{2020QuantumDutta}. At high temperatures, spatially correlated fluctuation was found to suppress the transition from coherent to incoherent dynamics\cite{2010Coherentprb}. In a loop system, spatially correlated fluctuation was shown to differentiate the best energy transfer pathway and obtain an optimal trapping probability\cite{2010QuantumTuning}. In terms of density matrix dynamics, simulations were run on models with spatially uncorrelated and correlated fluctuations, and the results revealed that spatially correlated fluctuations had no effect on single-exciton dynamics, whereas spatially uncorrelated transition energy fluctuations damped interband coherence\cite{2011Excitondy}. Some studies in the field of 2D electronic spectra\cite{2017Signatures} discovered that increasing the spatially correlated fluctuations increased the lifetime of oscillatory 2D signals at rephasing cross-peaks and non-rephasing diagonal peaks, while decreasing it at rephasing diagonal peaks and non-rephasing cross-peaks. Bath fluctuations can affect the excitation energy transfer dynamics even in the weak system-bath coupling regime by changing the spectral overlap factor between different chromophores\cite{2009Sharedmore,PhysRevA.84.053818,Scully15097,2014Noise}. Cao and colleagues\cite{doi:10.1021/jp4047243} demonstrated how spatial correlations produce similar results in an efficient energy transfer process for a multi-chromophoric system. In addition to the previously mentioned research efforts assessing the effects of spatially correlated fluctuations on quantum coherence and exciton dynamics, it was used to understand incoherent processes such as electrical transport in polymer films\cite{PhysRevB.63.245326} and fluorescent resonant energy transfer in one-dimensional polyproline peptides\cite{2007Fluorescentyu}.

Because of the prospect of efficient photoelectric conversion efficiency, some studies have been drawn to the quantum photovoltaic behavior in the quantum dot (QD) photocell in the last decade\cite{2020High,2010Quantum,Dorfman2746,Zhao2019,2011Peak,2013Optimal,ZHONG2021104094}. The multi-band QD photocell\cite{Zhao2019} scheme demonstrated a higher current output due to more absorbed low-energy photons, resulting in a higher output efficiency. The electron tunneling effect between two QDs has been shown to result in the redistribution of populations on two QDs in a DQD photocell \cite{ZHONG2021104094}, leading to an improvement in photovoltaic properties. Furthermore, quantum coherence has been shown to be advantageous in semiconductor QDs \cite{2007Analytical} and heterostructures \cite{PhysRevA.63.053803,2014Noise}.

Quantum coherence, on the other hand, has been shown to modify photon absorption and emission\cite{2010Quantum,Dorfman2746} in atomic systems, including lasing without inversion\cite{1996Electromagnetically}, electromagnetically induced transparency\cite{2009Plasmonic,0Light}. As a result, noise-induced quantum interference was demonstrated to drive photo-carrier generation dynamics at polymeric semiconductor heterojunctions\cite{2014Noise}, and it was thought to be an effective method of improving photovoltaic performance\cite{PhysRevA.84.053818,Scully15097}. To the best of our knowledge, the spatially correlated fluctuations couples to the conduction bands (CBs) in a DQD photocell scheme have received less attention in all previous studies.

It is thus intriguing to investigate how it affects a DQD photocell scheme when spatially correlated fluctuation is considered. In a DQD photocell model with two CBs coupled by spatially correlated fluctuations, we evaluate photovoltaic performance using photovoltaic current and output efficiency. The following is how the paper is structured. In Section II, we define notation and introduce the Hamiltonian for the DQD photocell model. The photovoltaic DQD cell's dynamics behavior in terms of reduced density matrix dynamics is presented in Sec.III. Sec.IV describes the delayed response of quantum fluctuation on photovoltaic properties. Sec.V discusses strategies for improving photovoltaic performance and suppressing delayed response. In Section VI, we include and make some remarks.

\section{Hamiltonian for the DQD photocell model}

\begin{figure}[htp]
\center
\subfigure[]{\includegraphics[width=0.5\columnwidth]{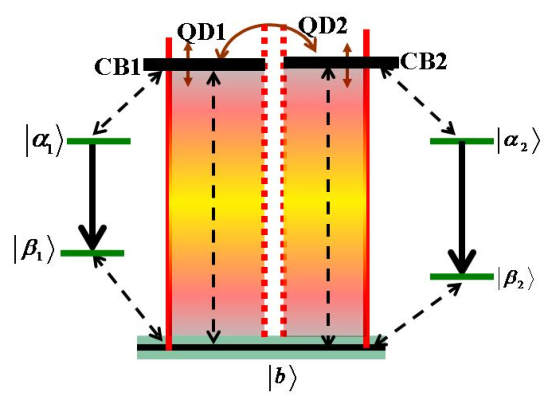}}
\hspace{0.10in}
\subfigure[]{\includegraphics[width=0.5\columnwidth]{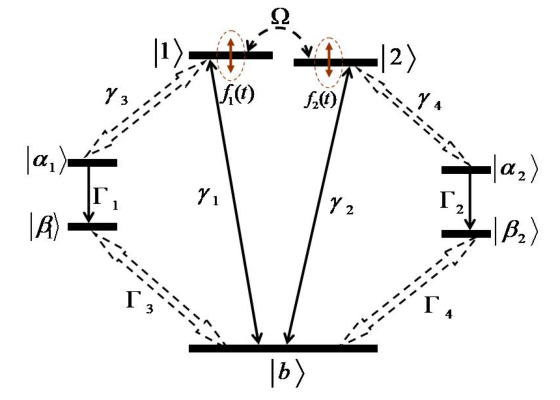}}
\caption{(Color online)(a) Schematic of electronic states of the DQD in both their eigenstate basis (solid lines). Solar radiation (depicted by warm-toned zones) drives transitions between the valence band (VB) and conduction band (CB) formed the photocell p-n junction. Other transitions are driven by ambient thermal phonon reservoir. Charge transfer to the left/right lead (denoted by states $|\alpha_{1}\rangle$ and $|\alpha_{2}\rangle$) are illustrated by dashed double-headed arrows. States $|\alpha_{i}\rangle$ and $|\beta_{i}\rangle$($_{i=1,2}$) connected to loads form the output electric current. (b) Corresponding energy-level diagram of the DQD photocell model.}
\label{Fig.1}
\end{figure}

The discovery of intermediate band solar cell (IBSC) has made a breakthrough in the field of modern photovoltaic research\cite{1961Shockley,1997Luque}, which was first introduced by Marti with the efficiency 63.2\(\%\) exceeding the previous maximum 33\(\%\) efficiency proposed by Shockley and Queisser. However, it is difficult to achieve in practical models due to manufacturing constraints. An intermediate band (IB£©can be achieved by the QDs\cite{2011Yang,2014Aly} due to their multifunctionality such as small size and tuning the range of band. The formed IB provides easier transportation path for the carriers. At the same time, it may also promote the electrons to the conduction band (CB) increasing the overall generation rate\cite{2014Aly,2017Jian}. This kind of structure may result in the multistep absorption process when arranging them in parallel or series to couple them either by tunneling or capacitively. Hence, double quantum dots (DQDs) have emerged as versatile and efficient scaffolds to absorb light and then convert optical energy into other useful forms of energy. The wave functions of closely spaced QDs get coupled and form minibands. These minibands exhibit new optoelectronics properties which is quite unique from the parent semiconductor materials\cite{2011Sugaya}. Different from these previous work, we focus on the photovoltaic performance in a DQD photocell with spatially correlated fluctuation.

We propose a DQD system with each dot contacting its individual electron reservoir at zero temperature and zero bias in this suggested DQD photocell model, and the system interacts with ambient radiation and phonon thermal reservoirs (See Fig.1~(a)). In this DQD photocell model (designated as QD1 and QD2 in Fig.1~(a), the level $|b\rangle$ represents the valence band (VB), while levels $|1\rangle$ and $|2\rangle$  represent the conduction bands (CB1 and CB2). Thermal solar photons drive $|b\rangle$ $\leftrightarrow$  $|i\rangle$ transitions and have average occupation numbers $n_{i}(\frac{E_{ib}}{K_BT_s})$=$[exp(\frac{E_{ib}}{K_BT_s})-1]^{-1}$  with the energy gap $E_{ib}$($_{i=1,2}$). At temperature $T_{a}$, low-energy transitions $|i\rangle$ $\leftrightarrow$ $|\alpha_{i}$ and  $|\beta_{i}\rangle$ $\leftrightarrow$ $|b\rangle_{(i=1,2)}$  are driven by ambient thermal phonons. The corresponding phonon occupation numbers are $n_{3,4} ={[exp(\frac{E_{i\alpha_i}}{K_BT_a})-1]}^{-1}$ and $n_{5,6} ={[exp(\frac{E_{\beta_{i}b}}{K_BT_a})-1]}^{-1}$ for each. The electric charges were then transferred to the external loads through the transitions $|\alpha_{i}\rangle$ $\leftrightarrow$ $|\beta_{i}\rangle$.

Unlike the earlier DQD photocell model\cite{ZHONG2021104094,ZHONG2021104503}, the quantum fluctuations coupling to the CBs (corresponding to the levels $|1\rangle$ and $|2\rangle$ in Fig.1~(b)) will be examined due to its interaction with the surrounding environment. In this suggested photocell model, we will investigate spatially correlated stochastic noise processes and explain how the spatial correlation influences charge transfer. As a result, the total Hamiltonian of the DQD photovoltaic model may be stated as (For convenience, we set $\hbar=1$),

\begin{align}
 \hat{H_{T}}=\hat{H}_{s}+\hat{H}_{P}+\hat{H}_{s-P},    \label{eq1}
\end{align}

\noindent Without loss of generality, the DQD system has three states: the left dot occupied state $|1\rangle$, the corresponding right one $|2\rangle$, and the ground state $|b\rangle$ with both dots empty owing to the strong Coulomb repulsion limit(see in Fig.\ref{Fig.1}(b))\cite{RevModPhys.75.1}. Hence, the Hamiltonian $\hat{H}_{s}$ in Eq.(\ref{eq1}) describes these states with an extra electron in the left or right dot (i.e. in the CB states $|1\rangle$ or $|2\rangle$ ) and it contains the following physical elements,

\begin{align}
\hat{H}_{s}=\hat{H}_{D}+\hat{H}_{L}+\hat{H}_{L-D},   \label{eq2}
\end{align}
\noindent in which $\hat{H}_{D}$ depicts the Hamiltonian of the DQD system with quantum fluctuation taken into account,

\begin{align}
 \hat{H}_{D}=\frac{\varepsilon}{2}\hat{\sigma}_{z}+\Omega\hat{\sigma}_{x}+\sum_{i=1,2}f_{i}(t)\hat{d}_{i}^ {\dag}\hat{d}_{i},  \label{eq3}
\end{align}

In the first term of the above Eq.(\ref{eq3}), $\varepsilon$ describes the electrostatic energy difference between the two CB states, and the Pauli matrix is defined as\cite{RevModPhys.75.1} $\hat{\sigma}_{z}$ = $\hat{d}_{1}^ {\dag}\hat{d}_{1}$ - $\hat{d}_{2}^ {\dag}\hat{d}_{2}$  with the subspace of states $\hat{d}_{i}^ {\dag}$ ($\hat{d}_{i}$)$_{(i=1,2)}$ representing the creation (annihilation ) operators and an electron in the left or right quantum dot, respectively.  The second term in above equation depicts the tunnel effect of an electron between the two dots with $\Omega$ being the tunnel matrix element of an electron, and with Pauli operator $\hat{\sigma}_{x}$ = $\hat{d}_{1}^ {\dag}$$\hat{d}_{2}$ + $\hat{d}_{2}^ {\dag}$$\hat{d}_{1}$. If a bias DC voltage is applied to this DQD system, the two eigenstates will be splitted by $\omega$: $\hbar\omega=\sqrt{\varepsilon^{2}+\Omega^{2}}$\cite{RevModPhys.75.1}. The last item in Eq.(\ref{eq3}) describes the spatially correlated fluctuation coupling to the two CB states owing to their interactions with the ambient environment. $f_{i}(t)$ is the fluctuating frequency on the CB states (corresponding to the levels $|1\rangle$ and $|2\rangle$ in Fig.\ref{Fig.1}(b)) and is considered as a stochastic process with the average of frequency fluctuation $ \langle f_{i}(t)\rangle $= 0 for each state. In this model, we will consider the fluctuation generated by the spatially correlated stochastic noise, and show what impact it will have and how to optimize the parameters of the DQD system in different ambient environment to control quantum photovoltaic properties. Therefore, for the sake of convenience, the correlation function $ \langle f_{i}(0)f_{j}(t)\rangle_{(i,j=1,2)} $ is described in a exponential decay process\cite{PhysRevE.83.011906,npjUchiyama2018} as follows,

\begin{align}
\langle f_{i}(0)f_{j}(t)\rangle_{(i,j=1,2)} =k A^{2} exp[-\frac{t}{\tau_{ij}}],  \label{eq4}
\end{align}

\noindent where $k$ is defined over the range by $-1\leq $ k $ \leq 1$, its positive and negative values represent the noise with spatial or anti-spatial correlations. The extremal value $-1$ ($+1$) corresponds to perfectly anti-correlated (correlated) noise, respectively. $A$ describes the amplitude of quantum fluctuation, and $\tau_{1,2}$ is the correlation time of the fluctuation. Therefore, insights into the spatio-correlation and intensity of quantum fluctuations will be measured by the two mentioned parameters for the optimizing charge transfer and output efficiency in this DQD photovoltaic model,

In Eq.(\ref{eq2}), $\hat{H}_{L}$ depicts the Hamiltonian of the external electrodes attached to the DQD system, the two eletrodes are modeled as collection of noninteracting electron with energy $\hbar\omega_{ik}$  and momentum $ k $ in the \emph{i}th (\emph{i}=1,2) electrode as follows,

\begin{align}
\hat{H}_{L}=\sum_{i=1,2}\sum_{k}\omega_{ik}\hat{a}_{ik}^ {\dag}\hat{a}_{ik},    \label{eq5}
\end{align}

\noindent in which  $\hat{a}_{ik}^ {\dag}$ ($\hat{a}_{ik}$) depicts the creation (annihilation) operator for an electron with momentum $ k $ in the \emph{i}th (\emph{i}=1,2) electrode. By making the rotating wave approximation approximation and with the Weisskopf-Wigner theory\cite{Zhao2019}, the Hamiltonian between the dot \emph{i} and lead \emph{i}(\emph{i}=1,2) with the corresponding coupling constant $ g_{ik}$ in Eq.(\ref{eq2}) can be given as,
\begin{align}
\hat{H}_{L-D}=\sum_{i=1,2}\sum_{k}g_{ik}\hat{d}_{ik}^ {\dag}\hat{a}_{ik} + H.c ,   \label{eq6}
\end{align}

\noindent where $H.c$ denotes the Hermite conjugate of the previous parts in Eq.(\ref{eq6}). In Eq.(\ref{eq1}), the ambient environment's  Hamiltonian is written as $\hat{H}_{P}=\sum_{q}\omega_{q}\hat{b}_{q}^ {\dag}\hat{b}_{q}$ with its \emph{q}-th noninteracting photon mode frequency \(\omega_{q}\). The interaction between the DQD photovoltaic system and photon degrees of freedom,  $\hat{H}_{s-P}$  in the Eq.(\ref{eq1}) is described by the Hamiltonian \cite{Wertnik2018Optimizing} with the corresponding coupling strength $\lambda_{q}$ as follows,

\begin{align}
\hat{H}_{s-P}=\hat{\sigma}_{x}\sum_{q}\lambda_{q}(\hat{b}_{q}^ {\dag}+\hat{b}_{q}),   \label{eq7}
\end{align}

In the current work, we devote to the influence generated by the quantum fluctuations which is the fundamental difference with the previous DQD photovoltaic model\cite{ZHONG2021104094,PhysRevB.87.035429,ZHONG2021104503}. In this DQD photovoltaic model, utilizing the Lindblad-like master equation, we analytically find that there is a decay response both to the dynamics of charge transfer and delivery efficiency, and the anti-spatial correlation fluctuation in a hotter thermal ambient environment brings out the damped photovoltaic properties owing to the damaged condition for noise-induced coherence. Comparing with the photovoltaic power enhanced by Fano-induced coherence\cite{PhysRevA.84.053818}, this work will reveal some different physical regimes on the efficient photovoltaic properties due to the quantum coherence deteriorated by the quantum fluctuations.

\section{Dynamics behavior of the photovoltaic DQD cell}

In the Schr$\ddot{o}$dinger picture the Born-Markov second order master equation takes the form,

\begin{widetext}
\begin{eqnarray}
\frac{d\hat{\rho}_{s}}{dt}=-i[\hat{H}_{s},\hat{\rho}_{s}]-\int^{\infty}_{0}d\tau\langle [\hat{H}_{s-P}(t),[\hat{H}_{s-P}(\tau),\hat{\rho}_{s}(t)\bigotimes\hat{\rho}_{P}(t_{0})]]\rangle_{\hat{H}_{P}} ,   \label{eq8}
\end{eqnarray}
\end{widetext}

\noindent where $\hat{\alpha}(\tau)$=$\exp [\frac{i}{\hbar}\hat{H}_{s}t] \hat{\alpha}(t)\exp [-\frac{i}{\hbar}\hat{H}_{s}t]$ is an operator $\hat{\alpha}$ at time $\tau$ in the interaction picture. And the symbol ``$\langle \cdots\rangle_{\hat{H}_{P}}$" denotes the trace operation on photons reservoir. Utilizing the second order perturbation method\cite{QinA2016,ZHONG2021104503}, the dynamics of electronic system can be described by the following Lindblad-like master equation:

\begin{equation}
\frac{d\hat{\rho}_{s}}{dt}=-i[\hat{H}_{s},\hat{\rho}_{s}]+\hat{L}\hat{\rho}_{s},   \label{eq9}
\end{equation}

\noindent The first term on the right-hand side of Eq. (\ref{eq9}) describes the unitary evolution of the DQD system and the second term accounts for the dissipative processes, and in which the superoperator $\hat{L}\hat{\rho}_{s}$ can be decomposed by,

\begin{eqnarray}
\hat{L}\hat{\rho}_{s}=\hat{L_{h}}\hat{\rho}_{s}+\hat{L_{c(1)}}\hat{\rho}_{s}+\hat{L_{c(2)}}\hat{\rho}_{s}+\hat{L_{rel}}\hat{\rho}_{s}, \label{eq10}
\end{eqnarray}

\noindent in which $\hat{L_{h}}\hat{\rho}_{s}, \hat{L_{c(1)}}\hat{\rho}_{s}, \hat{L_{c(2)}}\hat{\rho}_{s}, \hat{L_{rel}}\hat{\rho}_{s}$ are the superoperators related to the different dissipative processes between the DQD photovoltaic system and ambient photon (or phonon) reservoirs. As shown in Fig.\ref{Fig.1} (b), we assume that thermal solar photons are directed onto this DQD system and the solar photons are absorbed or radiated between the levels $|1\rangle$ ($|2\rangle$) and $|b\rangle$ at the rate $\gamma_{i}$. Hence, superoperator $\hat{L_{h}}\hat{\rho}_{s}$ describes transition process with the following expression,

\begin{align}
\hat{L_{h}}\hat{\rho}_{s} = \sum_{i=1,2}\frac{\gamma_{i}}{2}[ (n_{i}+1)\mathcal{D}[\hat{\sigma}_{bi}]\hat{\rho}_{s}+n_{i}\mathcal{D}[\hat{\sigma}_{bi}^{\dag}]\hat{\rho}_{s}], \label{eq11}
\end{align}

\noindent where $\hat{\sigma}_{bi} = |b\rangle\langle i|$ (i=1,2). Moreover, $n_{i} =[exp(x)-1]^{-1} $ is the average number of solar photons, $x=\frac{E_{i}b}{K_B T_s}_{(i=1,2)}$ with $T_{s}$, $E_{ib}$ being the solar's temperature and corresponding gap energy, respectively. The specific expression of the symbol $\mathcal{D}$ acting on any operator $\hat{A}$ is defined as $\mathcal{D}[\,\hat A\,]\hat\rho=2\hat A\hat\rho\hat A^{\dag}-\hat\rho\hat A^{\dag}\hat A-\hat A^{\dag}\hat A\hat\rho$. The dissipative transitions $|i_{(i=1,2)}\rangle$ $\leftrightarrow$ $|\alpha_{i(i=1,2)}\rangle$ is depicted by the super-operator $\hat{L_{c(1)}}\hat{\rho}_{s}$ with the following formula,

\begin{align}
\hat{L_{c(1)}}\hat{\rho}_{s}=\frac{\gamma_{3,4}}{2}[ (n_{3,4}+1)\mathcal{D}[\hat{\sigma}_{\alpha_ii}]\hat{\rho}_{s}+n_{3,4}\mathcal{D}[\hat{\sigma}_{\alpha_ii}^{\dag}]\hat{\rho}_{s}], \label{eq12}
\end{align}

\noindent where $\gamma_{3,4}$ are the spontaneous decay rates corresponding to the transitions $|1\rangle$ $\leftrightarrow$ $|\alpha_{1}\rangle$ and $|2\rangle$ $\leftrightarrow$ $|\alpha_{2}\rangle$, respectively.
And these transitions coupled to the ambient environment have the related phonon numbers $n_{3,4} = [exp(y)-1]^{-1}$, $y=\frac{E_{i\alpha_i}}{K_B T_a}_{(i=1,2)}$ with the ambient temperature $T_{a}$ and the energy difference $E_{i\alpha_i}$. The Pauli operator  $\hat{\sigma}_{\alpha_{i}i} $ is defined as $\hat{\sigma}_{\alpha_{i}i} = |\alpha_i\rangle\langle i|$ (i=1,2). Similarly,the dissipative process in the transition between $|\beta_{i(i=1,2)}\rangle$ $\leftrightarrow$ $|b\rangle$ can be written as

\begin{align}
\hat{L_{c(2)}}\hat{\rho}_{s}=\frac{\Gamma_{3,4}}{2}[ (n_{5,6}+1)\mathcal{D}[\hat{\sigma}_{\beta_ib}]\hat{\rho}_{s}+n_{5,6}\mathcal{D}[\hat{\sigma}_{\beta_ib}^{\dag}]\hat{\rho}_{s}], \label{eq13}
\end{align}

\noindent In these dissipation processes, $\Gamma_{3,4}$ are the spontaneous decay rate corresponding to the transitions  $|\beta_{1}\rangle$ $\leftrightarrow$ $|b\rangle$ and  $|\beta_{2}\rangle$ $\leftrightarrow$ $|b\rangle$(see in the fig.$\ref{Fig.1}$(b)), respectively. And the phonons numbers induced by the interference between these transitions are $n_{5,6} = {[exp(z)-1]}^{-1}$  with $z=\frac{E_{\beta_{i}b}}{K_B T_a}$ and $T_{a}$ being the ambient temperature, the energy difference $E_{\beta_{i}b}$$_{(i=1,2)}$. The Pauli operator $\hat{\sigma}_{\beta_{i}i} $ in Eq.(\ref{eq13}) is defined as $\hat{\sigma}_{\beta_{i}i} = |\beta_i\rangle\langle b|_{(i=1,2)}$ .

In this proposed DQD photovoltaic model, it contains two output terminals which can be directly connected to the external circuit. The photovoltaic currents are  proportional to the relaxation rates $\Gamma_{1,2}$ via
the dissipation processes $|\alpha_{i}\rangle$ $\leftrightarrow$ $|\beta_{i}\rangle$, which is defined as follows,

\begin{align}
\hat{L_{rel}}\hat{\rho}_{s} = \sum_{i=1,2}\frac{\Gamma_{i}}{2}(|\beta_i\rangle\langle \alpha_i|\hat\rho|\alpha_i\rangle\langle \beta_i|-|\alpha_i\rangle\langle \alpha_i|\hat\rho-\hat\rho|\alpha_i\rangle\langle \alpha_i|),\label{eq14}
\end{align}

\noindent Consequently, $|\alpha_{1}\rangle$ and $|\alpha_{2}\rangle$ are acted as charge separation states in this DQD photocell model. When the electrons are released from $|\alpha_{1}\rangle$ to $|\beta_{1}\rangle$ or from the second way $|\alpha_{2}\rangle$ to $|\beta_{2}\rangle$, the DQD is positively charged. The electrons are brought back to the neutral ground state $|b\rangle$ from $|\beta_{1}\rangle$ or $|\beta_{2}\rangle$, which completes a  closed loop. With this knowledge, this photovoltaic model is similar to the parallel circuit in classical electromagnetism. Thus, the collective currents can be defined by,

\begin{equation}
j=e\sum_{i=1,2}\Gamma_i{\rho}_{\alpha_{i}\alpha_{i}},   \label{eq15}
\end{equation}

Considering the Boltzmann distribution on the upper and lower levels $|\alpha_{i}\rangle$ and  $|\beta_{i}\rangle$ with the populations on the two levels being as follows,

\begin{align}
\rho_{\alpha_{i}\alpha_{i}}=& exp(-\frac{E_{\alpha_{i}}-\mu_{\alpha_{i}}}{k_{B}T_{a}}),\label{eq16}\\
\rho_{\beta_{i}\beta_{i}}=& exp(-\frac{E_{\beta_{i}}-\mu_{\beta_{i}}}{k_{B}T_{a}}), \label{eq17}
\end{align}

\noindent The total output voltage can be defined by the difference of the chemical potentials $\mu_{\alpha_{i}}$ and $\mu_{\beta_{i}}$ as follows,

\begin{equation}
eV =\sum_{i=1,2}(E_{\alpha_{i}}-E_{\beta_{i}}+K_{B}T_{a}\ln\frac{\rho_{\alpha_{i}\alpha_{i}}}\rho_{\beta_{i}\beta_{i}}),   \label{eq18}
\end{equation}

\noindent Utilizing the definitions of the current and the voltage mentioned in Eq.(\ref{eq15}) and Eq.(\ref{eq18}), we can intuitively deduce the output power of this DQD photovoltaic model with the classic definition $P=jV$.
Considering that incident solar photons are directed on the DQD photovoltaic system, the absorbed solar photons irrigate the DQD photovoltaic system with the input energy $P_{s}$. Therefore, the delivery efficiency of this photovoltaic system can be expressed as,

\begin{equation}
\eta=\frac{P}{Ps},                                 \label{eq19}
\end{equation}

Unlike the traditional processing method\cite{PhysRevB.87.035429,PhysRevA.84.053818}, we are not concerned with quantum coherence itself, but rather with the influence of quantum fluctuations on photovoltaic characteristics, and we attempt to fundamentally reveal the physical mechanism that affects quantum interference.
As a result, the influence generated by external environments is reduced during the interaction with the external environment under the Weisskopf-Wigner approximation\cite{1974On}, and the reduced matrix elements corresponding to Eg.(\ref{eq9}) are listed as follows:

\begin{widetext}\begin{eqnarray}
&\dot{\rho}_{11}=&-i\Omega(\rho_{21}-\rho_{12})-i\gamma_{3}(\rho_{\alpha_{1}1}-\rho_{1\alpha_{1}})-\gamma_{3}[(n_{3}+1)\rho_{11}-n_{3}\rho_{\alpha_{1}\alpha_{1}}]\nonumber\\&&-\gamma_{1}[(n_{1}+1)\rho_{11}-n_{1}\rho_{bb}],\nonumber\\
&\dot{\rho}_{22}=&i\Omega(\rho_{21}-\rho_{12})-i\gamma_{4}(\rho_{\alpha_{2}2}-\rho_{2\alpha_{2}})-\gamma_{4}[(n_{4}+1)\rho_{22}-n_{4}\rho_{\alpha_{2}\alpha_{2}}]\nonumber\\&&-\gamma_{2}[(n_{2}+1)\rho_{22}-n_{2}\rho_{bb}],\nonumber\\
&\dot{\rho}_{12}=&-i\varepsilon\rho_{12}-i\Omega(\rho_{22}-\rho_{11})-i\rho_{12}(f_{1}(t)-f_{2}(t))-\frac{\rho_{12}}{2}[\gamma_{1}(n_{1}+1)\nonumber\\&&+\gamma_{2}(n_{2}+1)+\gamma_{3}(n_{3}+1)+\gamma_{4}(n_{4}+1)],\nonumber\\
&\dot{\rho}_{21}=&i\varepsilon\rho_{21}+i\Omega(\rho_{22}-\rho_{11})+i\rho_{21}(f_{1}(t)-f_{2}(t))-\frac{\rho_{21}}{2}[\gamma_{1}(n_{1}+1)\nonumber\\&&+\gamma_{2}(n_{2}+1)+\gamma_{3}(n_{3}+1)+\gamma_{4}(n_{4}+1)],\nonumber\\
&\dot{\rho}_{\alpha_{1}\alpha_{1}}=&i\gamma_{3}(\rho_{\alpha_{1}1}-\rho_{1\alpha_{1}})+\gamma_{3}[(n_{3}+1)\rho_{11}-n_{3}\rho_{\alpha_{1}\alpha_{1}}]-\Gamma_{1}\rho_{\alpha_{1}\alpha_{1}},\nonumber\\
&\dot{\rho}_{\alpha_{2}\alpha_{2}}=&i\gamma_{4}(\rho_{\alpha_{2}2}-\rho_{2\alpha_{2}})+\gamma_{4}[(n_{4}+1)\rho_{22}-n_{4}\rho_{\alpha_{2}\alpha_{2}}]-\Gamma_{2}\rho_{\alpha_{2}\alpha_{2}},\nonumber\\
&\dot{\rho}_{\beta_{1}\beta_{1}}=&\Gamma_{1}\rho_{\alpha_{1}\alpha_{1}}-\Gamma_{3}[(n_{5}+1)\rho_{\beta_{1}\beta_{1}}-n_{5}\rho_{bb}],               \label{eq20}\\
&\dot{\rho}_{\beta_{2}\beta_{2}}=&\Gamma_{2}\rho_{\alpha_{2}\alpha_{2}}-\Gamma_{4}[(n_{6}+1)\rho_{\beta_{2}\beta_{2}}-n_{6}\rho_{bb}],\nonumber\\	&\dot{\rho}_{1\alpha_{1}}=&-i\gamma_{3}(\rho_{\alpha_{1}\alpha_{1}}-\rho_{11})+i(\omega_1-\frac{\varepsilon}{2}){\rho}_{1\alpha_{1}}-if_{1}(t){\rho}_{1\alpha_{1}}\nonumber\\&&-\frac{\rho_{1\alpha_1}}{2}[\gamma_{1}(n_{1}+1)+\gamma_{3}(2n_{3}+1)+\Gamma_1],\nonumber\\	&\dot{\rho}_{\alpha_{1}1}=&i\gamma_{3}(\rho_{\alpha_{1}\alpha_{1}}-\rho_{11})-i(\omega_1-\frac{\varepsilon}{2}){\rho}_{\alpha_{1}1}+if_{1}(t){\rho}_{\alpha_{1}1}\nonumber\\&&-\frac{\rho_{\alpha_{1}1}}{2}[\gamma_{1}(n_{1}+1)+\gamma_{3}(2n_{3}+1)+\Gamma_1],\nonumber\\ &\dot{\rho}_{2\alpha_{2}}=&-i\gamma_{4}(\rho_{\alpha_{2}\alpha_{2}}-\rho_{22})+i(\omega_2+\frac{\varepsilon}{2}){\rho}_{2\alpha_{2}}-if_{2}(t){\rho}_{2\alpha_{2}}\nonumber\\&&-\frac{\rho_{2\alpha_2}}{2}[\gamma_{2}(n_{2}+1)+\gamma_{4}(2n_{4}+1)+\Gamma_2],\nonumber\\	&\dot{\rho}_{\alpha_{2}2}=&i\gamma_{4}(\rho_{\alpha_{2}\alpha_{2}}-\rho_{22})-i(\omega_2+\frac{\varepsilon}{2}){\rho}_{\alpha_{2}2}+if_{2}(t){\rho}_{\alpha_{2}2}\nonumber\\&&-\frac{\rho_{\alpha_{2}2}}{2}[\gamma_{2}(n_{2}+1)+\gamma_{4}(2n_{4}+1)+\Gamma_2].\nonumber
\end{eqnarray}\end{widetext}

\section{Delayed response to the photovoltaic properties  due to the spatially correlated fluctuation }

To verify the photovoltaic properties in this DQD photocell model, we assess the charge transfer dynamics with some selected parameters, such as $E_{i}-E_{\alpha_{i}}$=$E_{\beta_{i}}-E_{b}=0.005eV$\cite{PhysRevA.84.053818}, $\gamma_{1}$ =$6.28\gamma$, $\gamma_{2}$ =$1.18\gamma$, $\gamma_{3}$=$\gamma_{4}$=$4.02\gamma$, $\Gamma_{1}$=$3.58\gamma$, $\Gamma_{2}$=$3.15\gamma$, $\Gamma_{3}$=$\Gamma_{4}$=$0.04\gamma$ with $\gamma$=$10^{-8}Hz$. Considering the fact that a small number of solar photons are absorbed at per unit time as well as weak-coupling between the DQD system and ambient environment, the photons with $\hbar\omega_{1}$=1.64 eV, $\hbar \omega_{2}$=1.6 eV are absorbed in the generated photovoltaic current process, and the numbers of photoelectron are set as $n_{1}$=30 and  $n_{2}$=15 populated on the CB states $|1\rangle$ and $|2\rangle$, respectively. The correlation time of the fluctuation between two CBs, $\tau_{12}$= 2.5 femtosecond. For the sake of applicability, we do not use gap energy as a controlling parameter, but utilize interdot gap energy difference $\varepsilon$ to revaluate the photovoltaic properties in our subsequent discussion.

\begin{figure}[htp]
\center
\includegraphics[width=0.5\columnwidth]{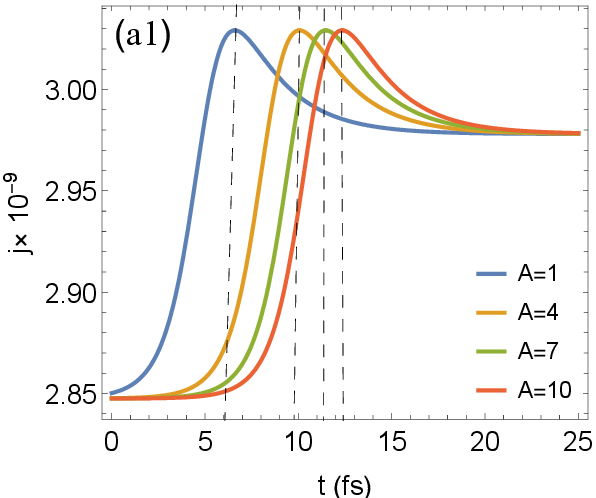}\hspace{0in}%
\includegraphics[width=0.5\columnwidth]{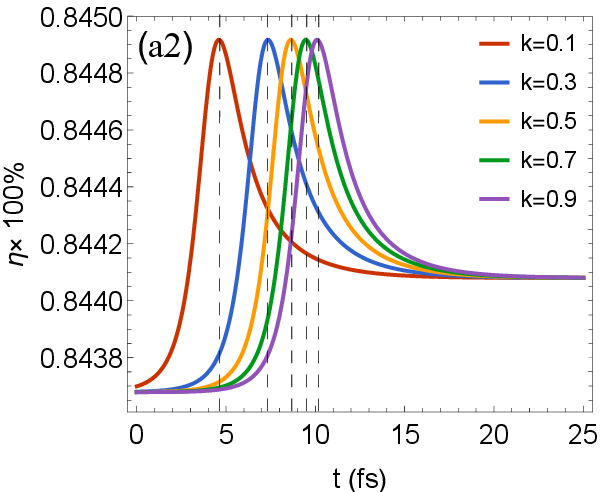}\hspace{0in}%
\caption{(Color online) Dynamics of the photovoltaic behavior influenced by spatially correlated fluctuations. (a1) Evolution of output current with different spatially correlated fluctuation amplitudes $A$, (a2) Evolution of delivery efficiency under different spatial correlation robustness $k$.}
\label{Fig.2}
\end{figure}

At the ambient temperature is close to room temperature, i.e., $T_{a}$=0.026$eV$, the interdot tunneling coefficient $\Omega$=0.03, gap energy difference $\varepsilon$=0.04$eV$, $k$=0.1 in Fig.\ref{Fig.2}(a1) while $\Omega$= 0.0025, $\varepsilon$=0.01$eV$, fluctuation amplitude $A$=0.9 in Fig.\ref{Fig.2}(a2) are utilized, respectively. We numerically solve the Eqs.(\ref{eq20}) and plot the evolution of output current with different spatially correlated fluctuation amplitudes $A$, and output efficiency with different spatio-correlation robustness $k$ in Fig.\ref{Fig.2}(a1) and (a2), respectively. In order to highlight the role of spatially correlated fluctuations in photovoltaic performance, other parameters, such as the self-structure and other performance parameters, ambient temperatures are not concluded in Fig.\ref{Fig.2}.

The dashed lines in Fig.\ref{Fig.2} (a1) and (a2) show a striking delayed response to the photovoltaic behavior, illustrating the prolonging peak times with the increments of $A$ and $k$ , respectively. The curves in Fig.\ref{Fig.2} (a1) show the delayed response as the spatially correlated fluctuation amplitudes $A$ increase, whereas the curves in Fig.\ref{Fig.2} (a2) show the increasing spatial correlation robustness $k$. Furthermore, we can see that the parameters $k$ and $k$ have no effect on the peaks and final steady values. However, we believe that the delayed response will affect the physical behavior of charge transfer and, as a result, the photovoltaic performance under various operating conditions. With this in mind, we continue to investigate the regulation influence on delayed response via some structural parameters, with the goal of clarifying the role of delayed response in this DQD photovoltaic model.

\section{Regulating strategies for delayed response}

\subsection{Delayed response regulated by the gap differences}

So far, the delayed response caused by spatially correlated fluctuations in the DQD photovoltaic cell has not been reported. We hypothesize that the delayed response caused by spatially correlated fluctuations is inhibited by a physical regime generated by the DQD itself. In order to better illustrate the underlying physical mechanisms, the spatial correlation robustness and fluctuation amplitudes are set to $k$=0.3, $A$=0.9, respectively, in the following discussion; all other parameters are the same as in Fig.\ref{Fig.2}.

\begin{figure}[htp]
\center
\includegraphics[width=0.5\columnwidth]{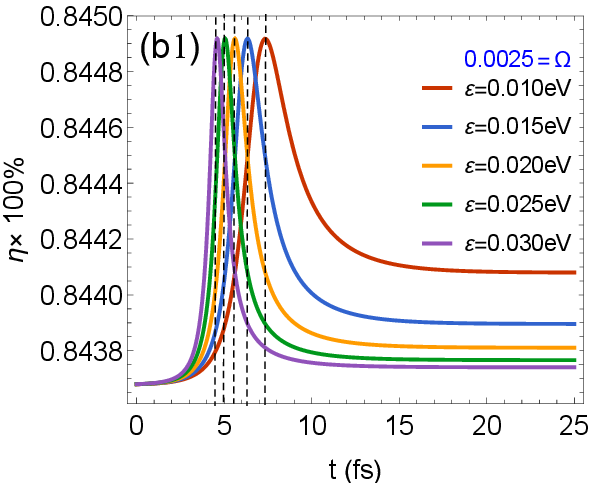}\includegraphics[width=0.5\columnwidth]{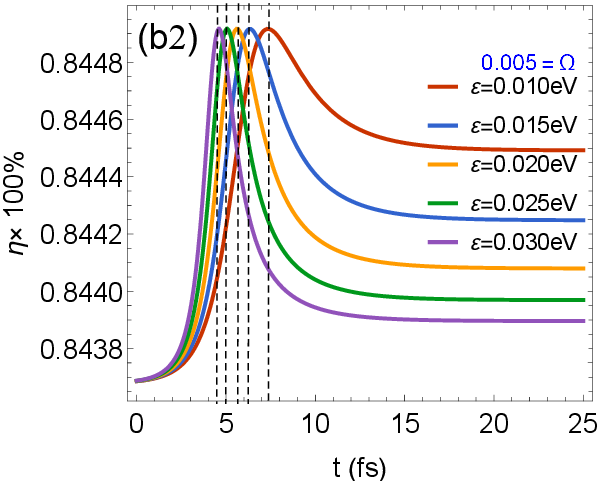}
\hspace{0in}%
\includegraphics[width=0.5\columnwidth]{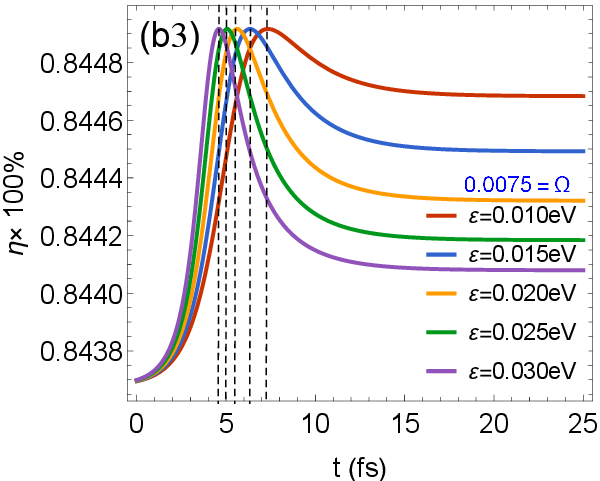}\includegraphics[width=0.5\columnwidth]{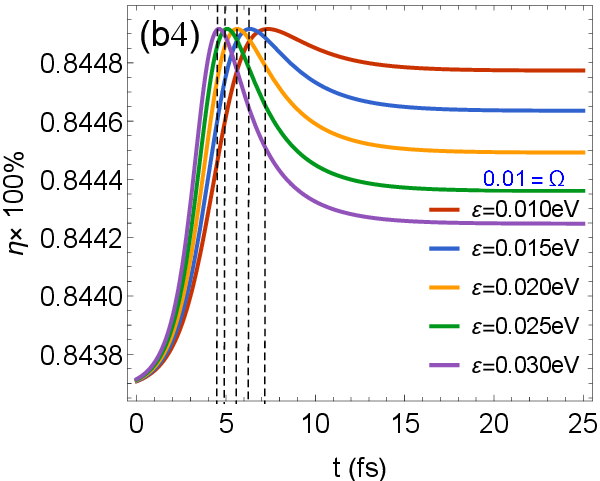}
\caption{(Color online) Evolution of the output photovoltaic efficiency controlled by gap energy difference with different inter-dot tunneling coefficients (b1) $\Omega$$=$0.0025,  (b2) $\Omega$$=$0.005, (b3) $\Omega$$=$0.0075, (b4) $\Omega$$=$0.01.}
\label{Fig.3}
\end{figure}

The curves in Fig.\ref{Fig.3} back up our theory about the underlying physical regime. The times corresponding to the peaks, as well as the final stable output efficiencies, decrease as the interdot gap energy difference $\varepsilon$ increases, as shown in Fig.\ref{Fig.3} (b1). In addition to these, two noteworthy conclusions emerge from a comparison of the curves in Fig.\ref{Fig.3} (b1) with (b2), (b3) and (b4). The peak efficiencies and delayed times, as shown by the peaks and their corresponding times, hardly change with increments of $\Omega$. The other is that stable efficiencies ascend in parallel with $\Omega$, as shown by the horizontal lines in Fig.3 from (b1) to (b4).

These results show that increasing the gap difference $\varepsilon$ can reduce the delayed response as well as the final steady output efficiencies, indicating that $\varepsilon$ describing the structure of the DQD may be an efficient approach to inhibiting the delayed response generated by the spatially correlated fluctuation. However, we hypothesize that the charge transfer between two adjacent quantum dots was reduced due to the increasing gap difference $\varepsilon$, resulting in a decrease in output efficiency, which could be the physical regime for the descending horizontal lines in Fig.\ref{Fig.3}. Meanwhile, the tunneling effect can reinforce charge transfer, but the spatially correlated fluctuation is not involved, resulting in more photoinduced charges being transported in the end while the time delayed response is almost unaffected. The tunneling coefficient $\Omega$ is well known to describe the energy transfer behavior between two dots, which can be flexibly tuned by applying an external bias voltage to the two dots\cite{RevModPhys.75.1,ZHAO2009}. As a result, in this proposed DQD photocell model, the applied bias voltage may be suggested to overcome the passive influence generated by the gap difference $\varepsilon$. Actually, these findings may explain why no significant quantum fluctuation effect has been observed in macroscopic DQD photovoltaic cells thus far.

\subsection{Delayed response dependent the spatial or anti-spatial correlations }

In Eq.(\ref{eq4}), the spatially correlated noise parameter $k$ with its positive or negative values represents the spatially or anti-spatially correlated degree, which directly reflects the correlated behaviors of quantum fluctuations owing to the environment noise. Therefore, it is possible to find some underlying physical strategies to restrain the delayed response by vesting insight into the spatially correlated noise parameter $k$ in different thermal environments. From Eq.(\ref{eq4}), a linear relationship can be observed between spatially correlated noise parameter $k$ and fluctuation intensity. Thus, the regulation features of noise parameter $k$ can directly reveal the physical behaviors of the spatially or anti-spatially correlated fluctuations.

\begin{figure}[htp]
\center
\includegraphics[width=0.5\columnwidth]{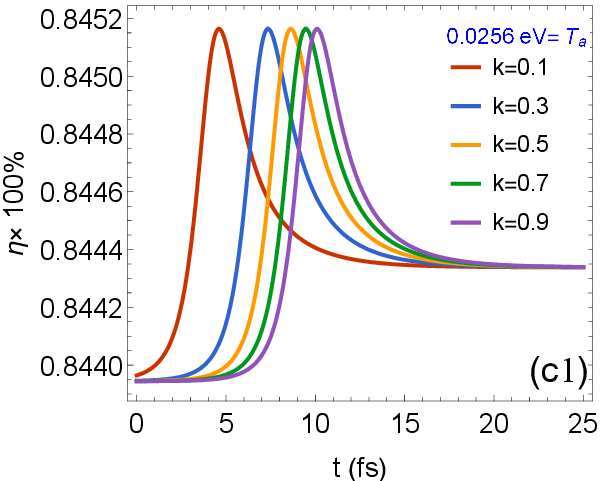}\includegraphics[width=0.5\columnwidth]{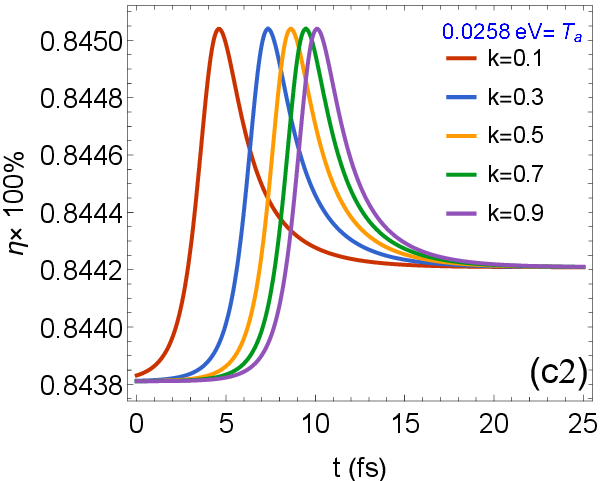}
\hspace{0in}%
\includegraphics[width=0.5\columnwidth]{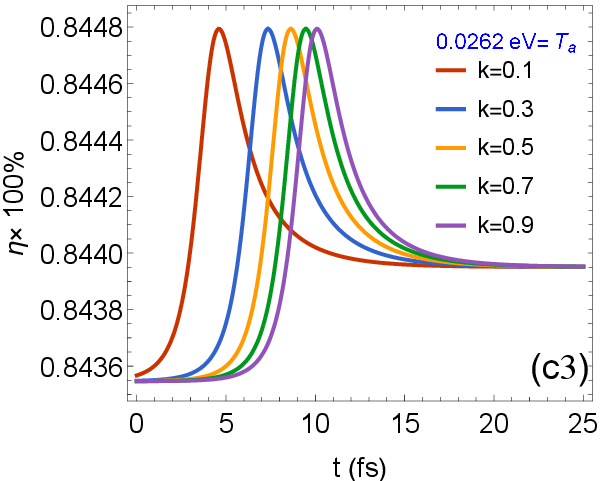}\includegraphics[width=0.5\columnwidth]{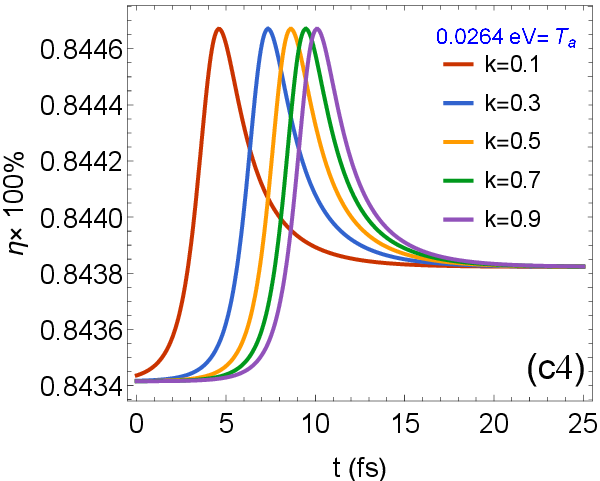}
\caption{(Color online) Evolution of the output photovoltaic efficiency with different spatial correlated noise parameters $k$ in different thermal ambient environment. (c1) $T_{a}$$=$0.0256 $eV$, (c2) $T_{a}$$=$0.0258 $eV$, (c3) $T_{a}$$=$0.0262 $eV$, (c4) $T_{a}$$=$0.0264 $eV$.}
\label{Fig.4}
\end{figure}

To facilitate the comparison with the results achieved in Fig.\ref{Fig.3}, the interdot tunneling coefficient $\Omega$ and gap difference $\varepsilon$ are set as $\Omega$$=$0.0025, $\varepsilon$=0.01 $eV$ in the upcoming discussion, and other parameters are the same to those in Fig.\ref{Fig.3}. Thereupon, the dependence of output photovoltaic efficiency on different spatially correlated noise parameter $k$ is plotted in Fig.\ref{Fig.4}. As the peaks of curves shown in Fig.\ref{Fig.4} (c1), the times corresponding the peak efficiencies illustrate that, the delayed response becomes more prominent and obvious with the increments of spatial correlations noise parameters $k$ at $T_{a}$$=$0.0256 $eV$. In addition to these, the peak photovoltaic efficiency slips from 84.52\(\%\), 84.50\(\%\), 84.48\(\%\) to 84.46\(\%\) when the ambient temperature increases in Fig.\ref{Fig.4} from (c1) to (c4) by 0.0256 $eV$, 0.0258 $eV$, 0.0262 $eV$ and 0.0264 $eV$, as well as their final steady values delivering efficiencies are uniform but descend with the increments of $T_{a}$.

The above results show that increasing the spatial correlations noise parameters $k$ results in a more prominent delayed response, but the DQD photovoltaic properties are diminished by the hotter thermal ambient environment.
Noise-induced coherence\cite{2014Noise} was shown in a decay model\cite{Scully15097} to break detailed balance and get more power out of a photocell due to noise-induced coherence against environmental decoherence. As a result, we believe that larger spatially correlated noise parameters $k$ only delayed but did not weaken noise-induced coherence. As a result, a more prominent delayed response is obtained, but peak efficiencies do not change as the spatially correlated fluctuation in the DQD photocell system increases. However, because noise-induced quantum coherence is impressionable to ambient temperatures\cite{Zhao2019}, the output photovoltaic performances decrease as ambient temperature increases.

\begin{figure}[htp]
\center
\includegraphics[width=0.5\columnwidth]{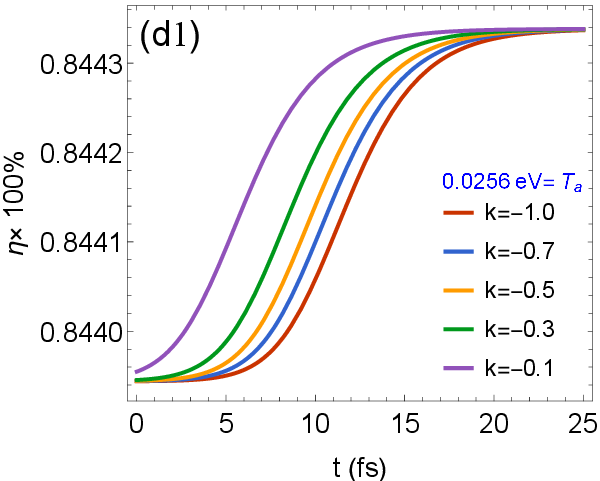}\includegraphics[width=0.5\columnwidth]{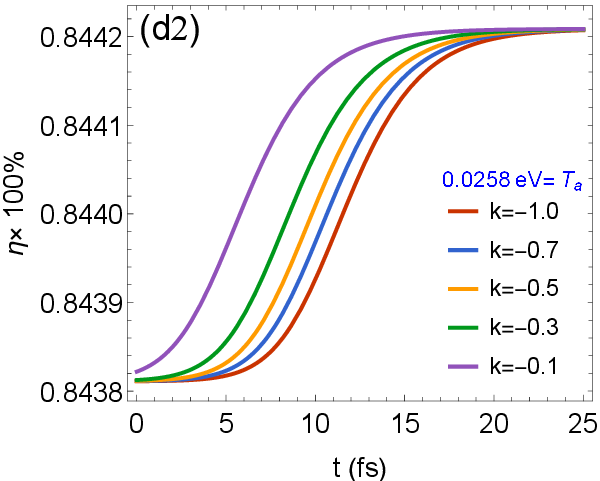}
\hspace{0in}%
\includegraphics[width=0.5\columnwidth]{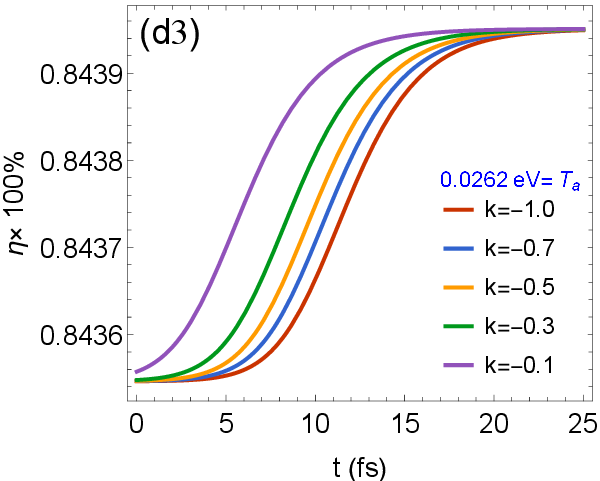}\includegraphics[width=0.5\columnwidth]{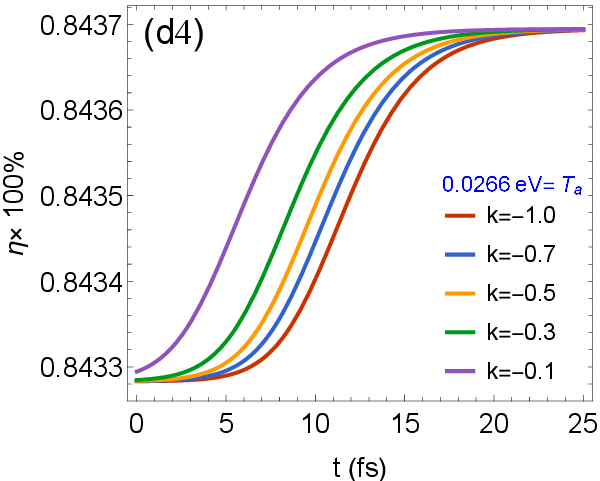}
\caption{(Color online) Evolution of photoelectric conversion efficiency with different  anti-spatial correlated noise parameters $k$  at different ambient temperatures. (d1) $T_{a}$$=$0.0256 $eV$, (d2) $T_{a}$$=$0.0258 $eV$, (d3) $T_{a}$$=$0.0262 $eV$, (d4) $T_{a}$$=$0.0266 $eV$.}
\label{Fig.5}
\end{figure}

As mentioned before, the anti-spatial correlated noise is represented by the negative values of the spatially correlated noise parameter $k$ \cite{npjUchiyama2018}. The negative property of anti-spatial correlated noise may show another physical property and bring out some different photovoltaic characteristics in this photocell model, which is our another concern. Different from the previous results with the spatial correlated fluctuation, a completely different linetype distributions, the hysteresis loop-style graph is obtained owing to the anti-spatially correlated noise in Fig.\ref{Fig.5}. As the curves shown in Fig.\ref{Fig.5}, the steady peak efficiencies decrease with the ambient temperature, which still suggests a negative effect on the steady output efficiencies due to $T_{a}$, as coincides with those shown in Fig.\ref{Fig.4}. Not only that, but the final steady peak efficiencies do not vary with the anti-spatial correlated noise parameters $k$, as well as the anti-spatial correlated noise parameters $k$ hardly exert influence over the delayed response (see the curves in Fig.\ref{Fig.5}(d1)). What's more, in the evolutionary process to the stable peaks, the photovoltaic efficiencies decrease with the increments of $k$ from -0.1, 0.3, 0.5, 0.7 to -1.0, i. e., the larger anti-spatial correlated noise parameters $k$, the lower output photovoltaic efficiency. So, results infer the output photovoltaic efficiency can be suppressed by the anti-spatial correlated noise parameters $k$.

Let's take a closer look at the physical implications implied by Fig.\ref{Fig.5}. When we compare the roles of spatial and anti-spatial correlated noise in Fig.\ref{Fig.4} and Fig.\ref{Fig.5}, we find that the spatially correlated noise causes a large time delay, but it has no effect on the output photovoltaic efficiency. However, anti-spatial correlated noise has no effect on peak photovoltaic efficiency, whereas in its evolutionary process, the stronger the anti-spatial correlated noise, the lower the output photovoltaic efficiency. Based on previous research on noise-induced coherence\cite{Scully15097,2014Noise}, we believe that the delayed photovoltaic properties are due to the generative process of noise-induced coherence being delayed by spatially correlated noise, whereas the reduced photovoltaic efficiency is due to the ambient temperature for noise-induced coherence being deteriorated by anti-spatial correlated noise. Furthermore, the hotter thermal ambient environment interferes with the generation of noise-induced coherence. As a result, the photoelectric conversion efficiency decreases as ambient temperature rises.

\section{Conclusions and remarks}

We introduce spatially correlated fluctuations to the two CBs in this proposed QDQ photovoltaic model and focus on the physical implications rather than the photovoltaic model itself. In this DQD photovoltaic model, the delayed response are demonstrated to the charge transfer and the process of ascending to the peak photovoltaic efficiency, while a natural adjustment feature was also verified on the delayed response in this DQD photocell model.
Further investigation reveals that the spatial correlated noise generates the delayed response, and the anti-spatial correlated noise can reduce the output photovoltaic efficiency, which is not the same as those in Ref.\cite{PhysRevE.78.050902}. Quantum coherence, on the other hand, has been shown to improve transport processes in both photosynthesis and photocells\cite{https://doi.org/10.1002/slct.201803554,doi:10.1021/acs.jpcb.0c10719,2019Intervalley}. Based on these, we hypothesize that the physical mechanism underlying the aforementioned conclusions is quantum coherence controlled by spatially correlated fluctuation. When the noise-induced coherence process is delayed by spatially correlated fluctuation, the output photovoltaic properties are inevitably delayed. While the anti-spatially correlated noise destroyed the condition for noise-induced coherence, the photovoltaic properties were reduced. Furthermore, the thermal ambient temperature degrades the environment for the generation of noise-induced coherence, which is one of the factors that reduces photovoltaic performance in this DQD photovoltaic model.

We'd like to make a few comments before we wrap up this paper. First, the spatially correlated fluctuations were imposed on the two CBs in our current work. However, the external load, i.e. the acceptor, may experience similar quantum fluctuations, and our work is unrelated to the results that the efficiencies are affected by acceptor fluctuations. However, the physical mechanism underlying such spatially correlated fluctuations and how it influences quantum photovoltaic properties remain unknown. On the other hand, because of the direct improvement in photovoltaic performance, revealing the relationship between spatially correlated fluctuations and long-lived coherence is a very important topic in DQD photovoltaic systems. As a result, we will broaden our current research to include DQD photovoltaic systems and the relationship between spatially correlated fluctuations and long-lived coherence. We believe that these relationships will help us understand the implicit physical significance of charge transfer and provide new insights into some fundamental problems in DQD photovoltaic systems.

\section*{Competing Interests}

The authors declare no competing financial or non-financial interests. This article does not contain any studies with human participants or animals performed by any of the authors. Informed consent was obtained from all individual participants included in the study.

\section*{Data Availability}
This manuscript has associated data in a data repository. [Authors¡¯ comment: All data included in this manuscript are available upon reasonable request by contacting with the corresponding author.]

\section*{Author contributions}

S. C. Zhao conceived the idea. S. N. Zhu performed the numerical computations and wrote the draft, and S. C. Zhao did the analysis and revised the paper. L. X. Xu and L. J. Chen participated in part of the discussion.

\section*{Acknowledgments}

This work is supported by the National Natural Science Foundation of China ( Grant Nos. 62065009 and 61565008 ), Yunnan Fundamental Research Projects, China( Grant No. 2016FB009 ).

\bibliography{reference}
\bibliographystyle{unsrt}
\end{document}